\documentclass[structabstract]{aa}  
\usepackage{graphicx}
\usepackage{natbib}
\usepackage{amssymb}
\usepackage{epstopdf}
\usepackage{multirow}
\def\thebibliography#1{\section*{References\markboth
 {REFERENCES}{REFERENCES}}\list
 {}{\setlength\labelwidth{1.4em}\leftmargin\labelwidth
 \setlength\parsep{0pt}\setlength\itemsep{0pt}
 \setlength{\itemindent}{-\leftmargin}
 \usecounter{enumi}}
 \def\newblock{\hskip .11em plus .33em minus -.07em}
 \sloppy
 \sfcode`\.=1000\relax}

\usepackage{txfonts}
\begin{document}
   \title{Detection of the hydroperoxyl radical HO$_2$ toward $\rho$ Oph A\thanks{Based on observations carried out with the IRAM 30m and the APEX telescopes. IRAM is supported by INSU/CNRS (France), MPG (Germany) and IGN (Spain). APEX is a collaboration between the Max Planck Institute for Radio Astronomy, the Onsala Space Observatory, and the European Southern Observatory.}}

   \subtitle{Additional constraints on the water chemical network}

   \author{B. Parise
          \inst{1}
          \and
          P. Bergman
          \inst{2}
          \and 
          F. Du 
          \inst{1}\thanks{Member of the International Max Planck Research School (IMPRS) 
          for Astronomy and Astrophysics at the Universities of Bonn and Cologne.}
          }

   \institute{Max-Planck-Institut f\"ur Radioastronomie, Auf dem H\"ugel 69, 53121 Bonn, Germany\\
              \email{bparise@mpifr-bonn.mpg.de}
         \and
   Onsala Space Observatory, Chalmers University of Technology, 439 92 Onsala, Sweden\\
             }

   \date{Received April 10, 2012 ; accepted April 25, 2012}

 
  \abstract
   {Hydrogen peroxide (HOOH) was recently detected toward $\rho$ Oph A. Subsequent astrochemical modeling that included reactions in the gas phase and on the surface of dust grains was able to explain the observed abundance, and highlighted the importance of grain chemistry in the formation of HOOH as an intermediate product in water formation. This study also predicted that the hydroperoxyl radical HO$_2$, the precursor of HOOH, should be detectable.}
   {We aim at detecting the hydroperoxyl radical HO$_2$ in $\rho$ Oph A.}
   {We used the IRAM 30m and the APEX telescopes to target the brightest HO$_2$ lines at about  130 and 260 GHz.  }
   {We detect five lines of HO$_2$ (comprising seven individual molecular transitions). The fractional abundance of HO$_2$ is found to be 
   about 10$^{-10}$, a value similar to the abundance of HOOH. This observational result is consistent with the prediction of the above 
   mentioned astrochemical model, and thereby validates our current understanding of the water formation on dust grains.  }
   {This detection, anticipated by a sophisticated gas-grain chemical model, demonstrates that models of grain chemistry have improved tremendously and that grain surface reactions now form a crucial part of the overall astrochemical network.}    


   \maketitle
%

\section{Introduction}

Water is an essential molecule in star-forming regions, in particular because it is the main constituent of the icy mantles of dust grains, one of the main repositories of oxygen, and in certain conditions an important gas coolant \citep[e.g. ][]{Nisini10}. It is moreover detected in virtually all star-forming environments in which it is searched for \citep[e.g. ][]{Caselli10, Hogerheijde11, vanDishoeck11}.  

Water is believed to form efficiently on the surface of dust grains through several different pathways. The relative importance of these pathways is not completely understood, however, and the reaction barriers involved are still under debate despite numerous recent laboratory experiments \citep{Ioppolo10,Cuppen10}. 
This is because isolating a particular reaction in the laboratory is extremely difficult, and the absolute measure of the reaction rates and barriers on the grains is almost impossible.

In this sense, astronomical observations of the precursors of water can be of great help to constrain the chemical network that leads to water formation. A significant step in the understanding of water formation was achieved thanks to the detection with APEX of HOOH in Oph A \citep{Bergman11b}. 
This detection has been carefully modeled with an astrochemical model combining reactions in the gas phase and on the grain surfaces based on the HME (Hybrid Moment Equation) method \citep{Du11}, applied to the conditions of the Oph A cloud \citep{Du12}. The model successfully reproduces simultaneously the abundances of the observed gas-phase HOOH, O$_2$, H$_2$CO and CH$_3$OH. The model also predicts that the hydroperoxyl radical (HO$_2$, or O$_2$H), the direct precursor leading to HOOH via hydrogenation, has a significant abundance, which makes its detection with current radio telescopes possible. Hydroperoxyl has not yet -- to our knowledge -- been detected in the interstellar medium in either its solid or its gaseous form.

We aim here at detecting this radical toward the SM1 core of the  $\rho$ Oph A cloud, where HOOH and O$_2$ \citep{Liseau12} were detected.
The paper is organized as follows. The observations are presented in Sect. 2, and the results in Sect. 3. We discuss the implications of this detection in Section 4.

\section{Observations}

\begin{figure}[!h]
\centering
\includegraphics[trim = 0 0 0 1.0cm, clip =true, width=9.0cm]{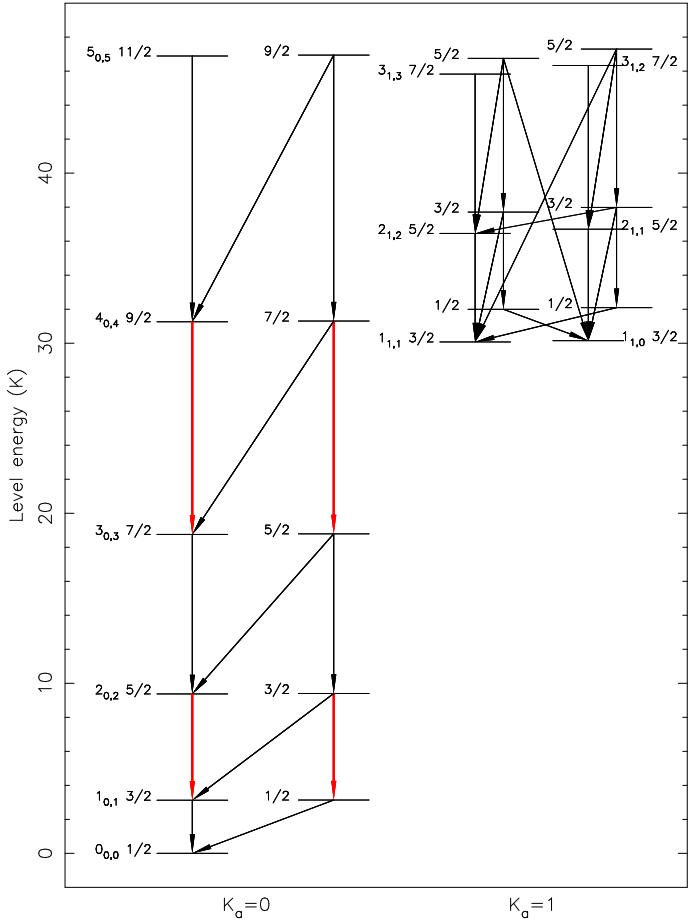}

\caption{Energy level diagram of HO$_2$ below 50 K. Only the $a$-type transitions listed in the JPL catalog were  included and are drawn as arrows. Red arrows indicate observed transitions.
The $b$-type transitions (connecting $K_a=0$ and 1 states) were omitted for the sake of clarity. Next to each level the quantum number designation is listed.}
\label{energy}
\end{figure}

The HO$_2$ radical is a light asymmetric rotor with an unpaired electronic spin. This causes each asymmetric rotor level (except $0_{0,0}$) to be split into a doublet of states. This is illustrated in the HO$_2$ energy level diagram in Fig. \ref{energy} where all states below 50 K are shown. It is based on the JPL catalog \citep{Pickett98}, which uses mm and submm spectroscopy data from \citet{Beers75}, \citet{Saito77}, and \citet{Charo82}. Also, the hyperfine splitting is large enough to be detectable. The spectrum of HO$_2$ exhibits both $a$- and $b$-type transitions with associated dipole moments $\mu_a=1.41$\,D and $\mu_b=1.54$\,D \citep{Saito80}.

We targeted lines of the HO$_2$ radical at 130.3\,GHz and 260.6\,GHz (see Table \ref{lines}), which were observed in the laboratory by \citet{Saito77} and \citet{Charo82}, respectively. We used the frequencies tabulated in the JPL database, which correspond to the measured values for these lines. The $1_{0,1}\to 0_{0,0}$  lines around 65 GHz are poorly suited for a ground-based search because they are heavily affected by the strong atmospheric absorption due to O$_2$. The $3_{0,3}\to 2_{0,2}$ lines (at 195\,GHz) fall in a frequency region where few receivers are currently in operation. They are also somewhat affected by the atmospheric absorption from the 183\,GHz water line. 
Likewise, the $5_{0,5}\to 4_{0,4}$ lines are affected by the atmospheric 325\,GHz water line. Hence, the targeted lines at 130 and 260 GHz are the best candidates for a detection in a low-excitation source (for which most of the population is likely to reside in the lower $K_a=0$ levels).

\begin{table*}[!ht]
\begin{center}
\caption{The HO$_2$ and SO$_2$ lines detected in the same setups.}
\label{lines}
\begin{tabular}{ccccccccccc}
\noalign{\smallskip}
\hline
\noalign{\smallskip}
Line & Frequency  & \multicolumn{3}{c}{Quantum numbers}     &   $A_{ul}$  & $E_{u}$ & $g_{u}$ & $\int T_{\mathrm{mb}} dv$ & $\Delta v$ & $v_{\rm lsr}$ \\
 & (GHz) &       Rotation &    $J_u \to J_l$   & $F_u \to F_l$  &   (s$^{-1}$)  & (K) &   & (mK\,s$^{-1}$) & (km\,s$^{-1}$)& (km\,s$^{-1}$) \\
\noalign{\smallskip}
\hline
\hline
\noalign{\smallskip}
1 & 130.26007  &  2$_{0, 2}$ $\rightarrow$ 1$_{0, 1}$  &  5/2$\rightarrow$3/2  &  3$\rightarrow$2  &  2.07$\times$10$^{-5}$  & 9.4 & 7 & 74$\pm$7 & 0.66$\pm$0.07 & 4.6$\pm$0.1 \\     
2 & 130.25813  &  2$_{0, 2}$ $\rightarrow$ 1$_{0, 1}$  &  5/2$\rightarrow$3/2  &  2$\rightarrow$1  &   1.85$\times$10$^{-5}$ &  9.4 & 5 & 54$\pm$8 & 0.65$\pm$0.11 & 4.6$\pm$0.1  \\    
3 & 130.46741  &  2$_{0, 2}$ $\rightarrow$ 1$_{0, 1}$  &  3/2$\rightarrow$1/2  &  2$\rightarrow$1  &  1.76$\times$10$^{-5}$  & 9.4 & 5 & 45$\pm$4 & 0.54$\pm$0.06 & 3.0$\pm$0.1  \\    
& 130.46368 & 2$_{0, 2}$ $\rightarrow$ 1$_{0, 1}$  &  3/2$\rightarrow$1/2  &  1$\rightarrow$0 &  1.16$\times$10$^{-5}$  & 9.4 & 3  &  $<$  26 \\ 
\noalign{\smallskip}
\hline
\noalign{\smallskip}
4a & 260.56614  &  4$_{0, 4}$ $\rightarrow$ 3$_{0, 3}$  &  9/2$\rightarrow$7/2   &  5$\rightarrow$4  &  1.86$\times$10$^{-4}$ & 31.3 & 11   & \multirow{2}{*}{\big\} 115$\pm$14} & \multirow{2}{*}{\big\} 0.88$\pm$0.13}& \multirow{2}{*}{--} \\
4b & 260.56566   &  4$_{0, 4}$ $\rightarrow$ 3$_{0, 3}$  &  9/2$\rightarrow$7/2    &  4$\rightarrow$3 & 1.81$\times$10$^{-4}$ & 31.3 &  9\\
5a & 260.77030  &  4$_{0, 4}$ $\rightarrow$ 3$_{0, 3}$  &  7/2$\rightarrow$5/2   &  4$\rightarrow$3  &   1.80$\times$10$^{-4}$ & 31.3 &  9 & \multirow{2}{*}{\big\} ~61$\pm$13} & \multirow{2}{*}{\big\} 0.91$\pm$0.24} & \multirow{2}{*}{--} \\
5b &   260.76967  &  4$_{0, 4}$ $\rightarrow$ 3$_{0, 3}$  &  7/2$\rightarrow$5/2&  3$\rightarrow$2  & 1.72$\times$10$^{-4}$ & 31.3 &  7\\
\noalign{\smallskip}
\hline
\noalign{\smallskip}
SO$_2$ &129.51481 & 10$_{2, 8}$ $\rightarrow$ 10$_{1, 9}$  &  &  & & & & 199$\pm$6 & 0.64$\pm$0.02 & 3.8$\pm$0.1 \\
SO$_2$ &131.01486  & 12$_{1, 11}$ $\rightarrow$ 12$_{0, 12}$  & & &  & & & 168$\pm$5 & 0.68$\pm$0.03 & 3.7$\pm$0.1\\ 
\noalign{\smallskip}
\hline
\noalign{\smallskip}
\end{tabular} 
\tablefoot{For lines 1 to 3 and SO$_2$, the frequencies are those measured in the laboratory and tabulated in the JPL database. The $v_{\rm lsr}$ of the lines are therefore computed based on these frequencies.  The hyperfine structure frequencies of lines 4 and 5 were computed from the energies tabulated in the JPL database (http://spec.jpl.nasa.gov/ftp/pub/catalog/archive/c033001.egy). Uncertainties on fluxes, linewidths and $v_{\rm lsr}$ are 1$\sigma$. The upper limit on the $ F = 1 \to 0$ line is the flux of the bump at the expected frequency (see Fig. \ref{fig_iram}).} 
\end{center}
\end{table*}

\begin{figure}[!h]
\centering
\includegraphics[width=7.2cm]{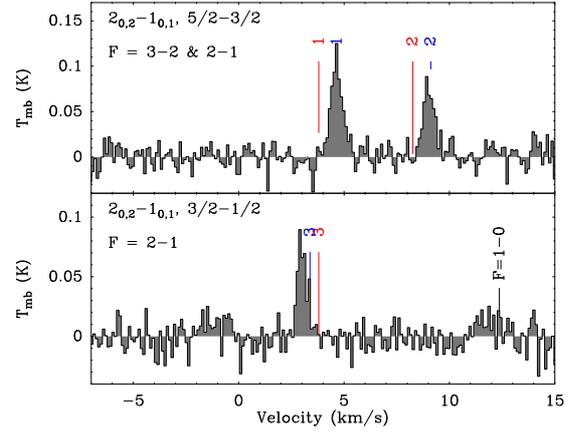}
\caption{Lines observed with the IRAM telescope. The red bars show the expected frequencies for the different lines, as measured in the laboratory \citep{Saito77}. The blue bars show the calculated frequencies from \citet{Charo82}. The black vertical line in the lower panel shows the position of the undetected $F = 1 \to 0$ line. }
\label{fig_iram}
\end{figure}

\begin{figure}[!h]
\centering
\includegraphics[width=7.2cm]{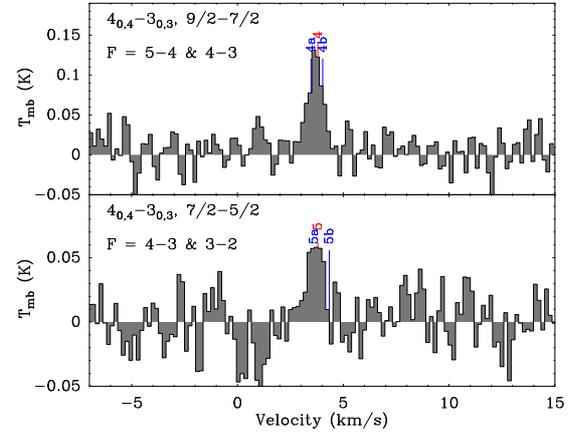}
\caption{Lines observed with the APEX telescope. The data were smoothed to a resolution of 0.18 km\,s$^{-1}$. The red bars show the expected frequencies for the different lines, as measured in the laboratory \citep{Charo82}. The blue bars show the computed hyperfine frequencies (see Table \ref{lines}). }
\label{fig_apex} 
\end{figure}

All observations targeted the SM1 position at $\alpha$(2000)=16$^h$26$^m$27\fs20 , $\delta$(2000)=$-$24$^\circ$24$'$04$''$. The IRAM 30m observations were carried out on Dec. 5, 2011, under average weather conditions with a precipitable water vapour (PWV) column of 5 mm.
The EMIR receiver  \citep{Carter12} was tuned in both polarizations to 130.3\,GHz. The angular resolution of the observations is 19$''$ at this frequency. 
The receiver was connected to FTS spectrometers,
which were used in their highest resolution mode, providing a 50\,kHz resolution. 
The focus was optimized on Saturn at the beginning of the observations. The pointing was checked locally every hour,
either on 1741-038, 1622-297, or 1730-130, depending on the time of the pointing, and was found to be better than 3\farcs8 (rms).
For the conversion to the $T_{\mathrm{mb}}$ scale, we adopt a main beam efficiency of 0.76, as extrapolated from the values given on the IRAM website\footnote{http://www.iram.es/IRAMES/mainWiki/Iram30mEfficiencies}.

The APEX observations were carried out  on Oct 10 and 16, 2011, under variable weather conditions (PWV\,=\,2.3 -- 3.7\,mm on Oct 10, PWV = 0.6
mm on Oct. 16). 
The APEX1 receiver \citep{Vassilev08} was tuned to 260.25\,GHz and connected to the XFFTS backend, leading to a velocity resolution of 0.09\,km\,s$^{-1}$. The angular resolution of the observations at this frequency is 24$''$. 
Because no planet was available during the observations, the focus was optimized on IRAS15194-5115 by maximizing the CO(2-1) line flux. The pointing was checked every hour by observing CO(2-1) on RAFGL1922, and was found to be always better than 3$''$.

\section{Results}

Fig. \ref{fig_iram} displays the three lines targeted with the IRAM telescope, and Fig. \ref{fig_apex} shows the two lines targeted with APEX.
Over the wide bandwidth observed with the FTS, no obvious other line is present, except for two SO$_2$ lines, which  allow us to confirm 
the $v_{\rm lsr}$ (3.8 km\,s$^{-1}$) and line width of the expected lines (see Table \ref{lines}). 

We detected three HO$_2$ lines with the IRAM telescope with a high signal-to-noise ratio, whose linewidths are consistent with the linewidth measured for the SO$_2$ lines, but whose frequencies  do not coincide exactly with those measured in the laboratory by \citet{Saito77}, shown as red vertical bars in Fig. \ref{fig_iram} (assuming a $v_{\rm lsr}$ of 3.8 km\,s$^{-1}$). 
\citet{Saito77} estimated an uncertainty on the laboratory frequency of 0.2\,MHz (i.e. 0.5 \,km\,s$^{-1}$ at 130\,\,GHz). \citet{Charo82} measured the frequency of other lines between 150 and 550\,GHz, and used the full dataset available at that time (including Saito's lines) to fit the HO$_2$ Hamiltonian. They found in particular a significant discrepancy (0.37\,MHz) between the frequency measured by \citet{Saito77} and the calculated frequency from the Hamiltonian for the two lines at 130.26\,GHz, which they excluded from their fit. The blue bars in Fig.\,\ref{fig_iram} show the calculated frequency from \citet{Charo82}. The first two detected lines coincide very closely with their theoretical frequencies. When adopting these new frequency values, the $v_{\rm lsr}$ of the lines become 
3.8  and 3.7 km s$^{-1}$, respectively, which is fully consistent with the source velocity as determined by SO$_2$ and several other molecules in the SM1 source \citep{Bergman11a}. 
The third line is still slightly offset, with a $v_{\rm lsr}$ of 3.4 km\,s$^{-1}$.

Because the source is relatively line-poor, we can assign the three lines to HO$_2$ with a high confidence level. An additional line (at 130.46368 GHz) is only tentatively detected, and we consider in the following only an upper limit on its flux. No blend with other plausible molecules was found at these frequencies in either the CDMS \citep{Muller01} or the JPL catalog \citep{Pickett98}. 

We detected two additional lines of HO$_2$ with the APEX telescope. Both lines have an unresolved hyperfine structure. A Gaussian fit of these two lines 
leads to a broader line width than for the other lines (see Table \ref{lines}), but this may be because of the hyperfine splitting, which for these lines corresponds to about 0.6\,--\,0.7\,km\,s$^{-1}$. Here again, no blend with any other plausible molecule is found.

In summary, we report here the detection of five different spectral features assigned to seven different lines of HO$_2$.

\section{Analysis and discussion}

\begin{figure}[!t]
\centering
\includegraphics[width=9cm]{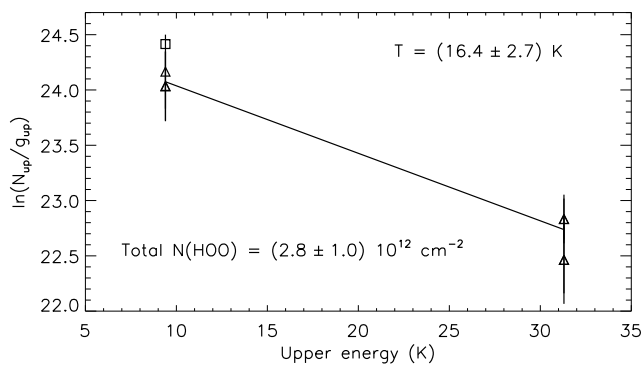}
\caption{HO$_2$ rotational diagram. A source size of 24$''$ is assumed. The triangles correspond to detected lines. The square corresponds to the
upper limit of the flux on the undetected line at 130.464 GHz. }
\label{rotdiag}
\end{figure}

We use the rotation diagram method to derive the abundance of HO$_2$, which assumes on the assumptions that the population distribution can be described by a single temperature, and that lines are optically thin. 
The rotation diagram obtained with the five detected spectral features is presented in Fig. \ref{rotdiag}. We assume for lines 4 and 5 that the
measured flux is coming from the two unresolved hyperfine components proportionally to $A_{ul}\times g_u$, as expected for two optically thin lines originating from two levels of the same energy. We assume a source size of 24$''$, as for hydrogen peroxide \citep{Bergman11b}, and as derived by the analysis of \citet{Bergman11a}. 
We derive a rotational temperature of 16\,$\pm$\,3\,K, slightly lower than the rotational temperature derived for HOOH \citep[22\,$\pm$\,3\,K,][]{Bergman11b}, but marginally consistent with it within the error bars. The derived HO$_2$ column density is (2.8 $\pm$ 1.0)\,$\times$\,10$^{12}$ cm$^{-2}$. This column density results in optically thin lines. Forcing the rotational temperature to be 22\,K as for HOOH, we derive a column density for HO$_2$ of 3.0\,$\times$\,10$^{12}$ cm$^{-2}$, very close to the column density derived with the fitted $T_{\rm rot}$ of 16.4\,K. 

The relatively high $A_{ul}$ of the 260 GHz lines (and the even higher spontaneous rates of the $b$-type transitions) indicate that the excitation could be somewhat subthermal for the density in SM1 \citep[$10^5 - 10^6$\,cm$^{-3}$,][]{Bergman11a}. This could imply that our derived column density is too high. 
To estimate this, we simply truncate the partition function such that the summation only takes place over $K_a=0$ states below 40 K and compare the result to the full partition function. We then find that the truncated part holds about 70\% of the level populations at $T=16\,{\mathrm K}$.
Therefore, in the unlikely case of extreme subthermal excitation (no population in $K_a > 0$ states and in $K_a = 0$ states above $4_{0,4}$), we may explain the observations by adopting a 30\% lower HO$_2$ column density. This lower column density falls within the errors and we conclude that subthermal excitation will have a minor impact on our derived column density.

From the H$_2$CO and CH$_3$OH analysis of the SM1 core, \citet{Bergman11a} determined an H$_2$ column density of 3\,$\times$\,10$^{22}$ cm$^{-2}$. The abundance of HO$_2$ is therefore $\sim$10$^{-10}$.

The gas-grain chemical model of \citet{Du12}, based on the detection of HOOH and other molecules, predicts that the HO$_2$\,/\,HOOH abundance ratio in $\rho$\,Oph A should be about three. In this model, at $T$\,$\sim$\,20 K, HO$_2$ is mainly formed on dust grains through the barrierless reaction O + OH at an early stage ($<$\,5\,$\times$\,10$^4$ yr), and through the reaction H + O$_2$ at a later
stage.
Here, we find observationally HO$_2$\,/\,HOOH\,$\sim$\,1, which is consistent with the model, taking into account the excitation uncertainty. This result is therefore an additional validation of our understanding of water formation on 
dust grains.

Motivated by our detection, we inspected previous spectral surveys for unidentified lines at the frequencies of the HO$_2$ molecule. No line is  
detected in the TIMASSS survey \citep{Caux11} toward the low-mass protostar IRAS16293$-$2422. \citet{Nummelin98} reported (in their Table 46) two unidentified lines in their Sgr B2 survey that might correspond to HO$_2$ frequencies, U260568 and U260771 toward Sgr B2(N) and U260569 toward Sgr B2(M). Because the lines have typical widths of 13\,km\,s$^{-1}$, line blending is a severe limiting factor for a secure line assignment. The Sgr B2 IRAM\,30\,m line survey coverage does not include the lines at 130 GHz (Belloche, \textit{ priv. comm.}). A 2mm unpublished survey from B. Turner made with the NRAO 12m telescope between 1993 and 1995 covers the two 130.26\,GHz lines \citep[but not the 130.46\,GHz lines,][]{Remijan08}. This doublet unfortunately falls 
very close to the $J$=3-2 SiO emission/absorption feature at 130268.7 MHz, making it very difficult to discern any weaker HO$_2 $ lines.
Assuming that the unidentified lines of the Sgr B2(N) survey by \citet{Nummelin98} are assigned to HO$_2$, and assuming $T_{\rm rot}$\,=\,50\,K \citep{Nummelin00}, we find $N$(HO$_2$)\,$\le$\,1.1\,$\times$\,10$^{14}$\,cm$^{-2}$, averaged in the 20$''$ beam. Using the H$_2$ column density derived by Nummelin et al. (3\,$\times$\,10$^{24}$\,cm$^{-2}$), 
this results in an upper limit on the fractional abundance of HO$_2$ of 4\,$\times$\,10$^{-11}$, somewhat below the detected abundance toward $\rho$~Oph A.

The result of this work is one of the rare examples of the detection of a new molecule after it was predicted by chemical modeling \citep{Du12}. Other examples include the detection of CF$^+$ \citep{Neufeld06}, motivated also by (gas-phase) model predictions. The novelty here is that the detection of HO$_2$ was predicted by models describing the chemistry on dust grain surfaces. This shows the increasing predictive power of these models, thanks to the numerical refinements developed to correctly treat the stochasticity of surface chemistry \citep[e.g.][]{Garrod09,Du11}, and the recent numerous laboratory experiments \citep[e.g.][]{Oba09, Ioppolo10, Cuppen10}. The era when dust chemistry could be described as ``the last refuge of the scoundrels" \citep{Charnley92} seems to be over. 

\begin{acknowledgements}
We thank the referee, P. Goldsmith, for comments that helped improving the manuscript.
We are grateful to the IRAM director, P. Cox, for granting us directorial time on the IRAM 30\,m telescope. We thank A. Gusdorf for carrying out the APEX observations. BP and FD are supported by the German Deutsche Forschungsgemeinschaft, DFG Emmy Noether project number PA1692/1-1.
\end{acknowledgements}

\bibliography{/Users/bparise/These/Manuscrit/biblio}

\begin{thebibliography}{27}
\expandafter\ifx\csname natexlab\endcsname\relax\def\natexlab#1{#1}\fi

\bibitem[{{Beers} \& {Howard}(1975)}]{Beers75}
{Beers}, Y. \& {Howard}, C.~J. 1975, \jcp, 63, 4212

\bibitem[{{Bergman} {et~al.}(2011{\natexlab{a}}){Bergman}, {Parise}, {Liseau},
  \& {Larsson}}]{Bergman11a}
{Bergman}, P., {Parise}, B., {Liseau}, R., \& {Larsson}, B. 2011{\natexlab{a}},
  \aap, 527, A39

\bibitem[{{Bergman} {et~al.}(2011{\natexlab{b}}){Bergman}, {Parise}, {Liseau},
  {Larsson}, {Olofsson}, {Menten}, \& {G{\"u}sten}}]{Bergman11b}
{Bergman}, P., {Parise}, B., {Liseau}, R., {et~al.} 2011{\natexlab{b}}, \aap,
  531, L8

\bibitem[{{Carter} {et~al.}(2012){Carter}, {Lazareff}, {Maier}, {Chenu},
  {Fontana}, {Bortolotti}, {Boucher}, {Navarrini}, {Blanchet}, {Greve}, {John},
  {Kramer}, {Morel}, {Navarro}, {Pe{\~n}alver}, {Schuster}, \&
  {Thum}}]{Carter12}
{Carter}, M., {Lazareff}, B., {Maier}, D., {et~al.} 2012, \aap, 538, A89

\bibitem[{{Caselli} {et~al.}(2010){Caselli}, {Keto}, {Pagani}, {Aikawa},
  {Y{\i}ld{\i}z}, {van der Tak}, {Tafalla}, {Bergin}, {Nisini}, {Codella}, {van
  Dishoeck}, {Bachiller}, {Baudry}, {Benedettini}, {Benz}, {Bjerkeli}, {Blake},
  {Bontemps}, {Braine}, {Bruderer}, {Cernicharo}, {Daniel}, {di Giorgio},
  {Dominik}, {Doty}, {Encrenaz}, {Fich}, {Fuente}, {Gaier}, {Giannini},
  {Goicoechea}, {de Graauw}, {Helmich}, {Herczeg}, {Herpin}, {Hogerheijde},
  {Jackson}, {Jacq}, {Javadi}, {Johnstone}, {J{\o}rgensen}, {Kester},
  {Kristensen}, {Laauwen}, {Larsson}, {Lis}, {Liseau}, {Luinge}, {Marseille},
  {McCoey}, {Megej}, {Melnick}, {Neufeld}, {Olberg}, {Parise}, {Pearson},
  {Plume}, {Risacher}, {Santiago-Garc{\'{\i}}a}, {Saraceno}, {Shipman},
  {Siegel}, {van Kempen}, {Visser}, {Wampfler}, \& {Wyrowski}}]{Caselli10}
{Caselli}, P., {Keto}, E., {Pagani}, L., {et~al.} 2010, \aap, 521, L29

\bibitem[{{Caux} {et~al.}(2011){Caux}, {Kahane}, {Castets}, {Coutens},
  {Ceccarelli}, {Bacmann}, {Bisschop}, {Bottinelli}, {Comito}, {Helmich},
  {Lefloch}, {Parise}, {Schilke}, {Tielens}, {van Dishoeck}, {Vastel},
  {Wakelam}, \& {Walters}}]{Caux11}
{Caux}, E., {Kahane}, C., {Castets}, A., {et~al.} 2011, \aap, 532, A23

\bibitem[{{Charnley} {et~al.}(1992){Charnley}, {Tielens}, \&
  {Millar}}]{Charnley92}
{Charnley}, S.~B., {Tielens}, A.~G.~G.~M., \& {Millar}, T.~J. 1992, \apjl, 399,
  L71

\bibitem[{{Charo} \& {de Lucia}(1982)}]{Charo82}
{Charo}, A. \& {de Lucia}, F.~C. 1982, Journal of Molecular Spectroscopy, 94,
  426

\bibitem[{{Cuppen} {et~al.}(2010){Cuppen}, {Ioppolo}, {Romanzin}, \&
  {Linnartz}}]{Cuppen10}
{Cuppen}, H.~M., {Ioppolo}, S., {Romanzin}, C., \& {Linnartz}, H. 2010,
  Physical Chemistry Chemical Physics (Incorporating Faraday Transactions), 12,
  12077

\bibitem[{{Du} \& {Parise}(2011)}]{Du11}
{Du}, F. \& {Parise}, B. 2011, \aap, 530, A131

\bibitem[{{Du} {et~al.}(2012){Du}, {Parise}, \& {Bergman}}]{Du12}
{Du}, F., {Parise}, B., \& {Bergman}, P. 2012, \aap, 538, A91

\bibitem[{{Garrod} {et~al.}(2009){Garrod}, {Vasyunin}, {Semenov}, {Wiebe}, \&
  {Henning}}]{Garrod09}
{Garrod}, R.~T., {Vasyunin}, A.~I., {Semenov}, D.~A., {Wiebe}, D.~S., \&
  {Henning}, T. 2009, \apjl, 700, L43

\bibitem[{{Hogerheijde} {et~al.}(2011){Hogerheijde}, {Bergin}, {Brinch},
  {Cleeves}, {Fogel}, {Blake}, {Dominik}, {Lis}, {Melnick}, {Neufeld},
  {Pani{\'c}}, {Pearson}, {Kristensen}, {Y{\i}ld{\i}z}, \& {van
  Dishoeck}}]{Hogerheijde11}
{Hogerheijde}, M.~R., {Bergin}, E.~A., {Brinch}, C., {et~al.} 2011, Science,
  334, 338

\bibitem[{{Ioppolo} {et~al.}(2010){Ioppolo}, {Cuppen}, {Romanzin}, {van
  Dishoeck}, \& {Linnartz}}]{Ioppolo10}
{Ioppolo}, S., {Cuppen}, H.~M., {Romanzin}, C., {van Dishoeck}, E.~F., \&
  {Linnartz}, H. 2010, Physical Chemistry Chemical Physics (Incorporating
  Faraday Transactions), 12, 12065

\bibitem[{{Liseau} {et~al.}(2012){Liseau}, {Goldsmith}, {Larsson}, {Pagani},
  {Bergman}, {Le Bourlot}, {Bell}, {Benz}, {Bergin}, {Bjerkeli}, {Black},
  {Bruderer}, {Caselli}, {Caux}, {Chen}, {de Luca}, {Encrenaz}, {Falgarone},
  {Gerin}, {Goicoechea}, {Hjalmarson}, {Hollenbach}, {Justtanont}, {Kaufman},
  {Le Petit}, {Li}, {Lis}, {Melnick}, {Nagy}, {Olofsson}, {Olofsson}, {Roueff},
  {Sandqvist}, {Snell}, {van der Tak}, {van Dishoeck}, {Vastel}, {Viti}, \&
  {Y$\backslash$ild$\backslash$iz}}]{Liseau12}
{Liseau}, R., {Goldsmith}, P.~F., {Larsson}, B., {et~al.} 2012, \aap, 541, A73

\bibitem[{{M{\" u}ller} {et~al.}(2001){M{\" u}ller}, {Thorwirth}, {Roth}, \&
  {Winnewisser}}]{Muller01}
{M{\" u}ller}, H.~S.~P., {Thorwirth}, S., {Roth}, D.~A., \& {Winnewisser}, G.
  2001, \aap, 370, L49

\bibitem[{{Neufeld} {et~al.}(2006){Neufeld}, {Schilke}, {Menten}, {Wolfire},
  {Black}, {Schuller}, {M{\"u}ller}, {Thorwirth}, {G{\"u}sten}, \&
  {Philipp}}]{Neufeld06}
{Neufeld}, D.~A., {Schilke}, P., {Menten}, K.~M., {et~al.} 2006, \aap, 454, L37

\bibitem[{{Nisini} {et~al.}(2010){Nisini}, {Benedettini}, {Codella},
  {Giannini}, {Liseau}, {Neufeld}, {Tafalla}, {van Dishoeck}, {Bachiller},
  {Baudry}, {Benz}, {Bergin}, {Bjerkeli}, {Blake}, {Bontemps}, {Braine},
  {Bruderer}, {Caselli}, {Cernicharo}, {Daniel}, {Encrenaz}, {di Giorgio},
  {Dominik}, {Doty}, {Fich}, {Fuente}, {Goicoechea}, {de Graauw}, {Helmich},
  {Herczeg}, {Herpin}, {Hogerheijde}, {Jacq}, {Johnstone}, {J{\o}rgensen},
  {Kaufman}, {Kristensen}, {Larsson}, {Lis}, {Marseille}, {McCoey}, {Melnick},
  {Olberg}, {Parise}, {Pearson}, {Plume}, {Risacher}, {Santiago}, {Saraceno},
  {Shipman}, {van Kempen}, {Visser}, {Viti}, {Wampfler}, {Wyrowski}, {van der
  Tak}, {Y{\i}ld{\i}z}, {Delforge}, {Desbat}, {Hatch}, {P{\'e}ron}, {Schieder},
  {Stern}, {Teyssier}, \& {Whyborn}}]{Nisini10}
{Nisini}, B., {Benedettini}, M., {Codella}, C., {et~al.} 2010, \aap, 518, L120

\bibitem[{{Nummelin} {et~al.}(1998){Nummelin}, {Bergman}, {Hjalmarson},
  {Friberg}, {Irvine}, {Millar}, {Ohishi}, \& {Saito}}]{Nummelin98}
{Nummelin}, A., {Bergman}, P., {Hjalmarson}, {\AA}., {et~al.} 1998, \apjs, 117,
  427

\bibitem[{{Nummelin} {et~al.}(2000){Nummelin}, {Bergman}, {Hjalmarson},
  {Friberg}, {Irvine}, {Millar}, {Ohishi}, \& {Saito}}]{Nummelin00}
{Nummelin}, A., {Bergman}, P., {Hjalmarson}, {\AA}., {et~al.} 2000, \apjs, 128,
  213

\bibitem[{{Oba} {et~al.}(2009){Oba}, {Miyauchi}, {Hidaka}, {Chigai},
  {Watanabe}, \& {Kouchi}}]{Oba09}
{Oba}, Y., {Miyauchi}, N., {Hidaka}, H., {et~al.} 2009, \apj, 701, 464

\bibitem[{{Pickett} {et~al.}(1998){Pickett}, {Poynter}, {Cohen}, {Delitsky},
  {Pearson}, \& {M{\"u}ller}}]{Pickett98}
{Pickett}, H.~M., {Poynter}, R.~L., {Cohen}, E.~A., {et~al.} 1998, \jqsrt, 60,
  883

\bibitem[{{Remijan} {et~al.}(2008){Remijan}, {Leigh}, {Markwick-Kemper}, \&
  {Turner}}]{Remijan08}
{Remijan}, A.~J., {Leigh}, D.~P., {Markwick-Kemper}, A.~J., \& {Turner}, B.~E.
  2008, ArXiv e-prints

\bibitem[{{Saito}(1977)}]{Saito77}
{Saito}, S. 1977, Journal of Molecular Spectroscopy, 65, 229

\bibitem[{{Saito} \& {Matsumura}(1980)}]{Saito80}
{Saito}, S. \& {Matsumura}, C. 1980, Journal of Molecular Spectroscopy, 80, 34

\bibitem[{{van Dishoeck} {et~al.}(2011){van Dishoeck}, {Kristensen}, {Benz},
  {Bergin}, {Caselli}, {Cernicharo}, {Herpin}, {Hogerheijde}, {Johnstone},
  {Liseau}, {Nisini}, {Shipman}, {Tafalla}, {van der Tak}, {Wyrowski},
  {Aikawa}, {Bachiller}, {Baudry}, {Benedettini}, {Bjerkeli}, {Blake},
  {Bontemps}, {Braine}, {Brinch}, {Bruderer}, {Chavarr{\'{\i}}a}, {Codella},
  {Daniel}, {de Graauw}, {Deul}, {di Giorgio}, {Dominik}, {Doty}, {Dubernet},
  {Encrenaz}, {Feuchtgruber}, {Fich}, {Frieswijk}, {Fuente}, {Giannini},
  {Goicoechea}, {Helmich}, {Herczeg}, {Jacq}, {J{\o}rgensen}, {Karska},
  {Kaufman}, {Keto}, {Larsson}, {Lefloch}, {Lis}, {Marseille}, {McCoey},
  {Melnick}, {Neufeld}, {Olberg}, {Pagani}, {Pani{\'c}}, {Parise}, {Pearson},
  {Plume}, {Risacher}, {Salter}, {Santiago-Garc{\'{\i}}a}, {Saraceno},
  {St{\"a}uber}, {van Kempen}, {Visser}, {Viti}, {Walmsley}, {Wampfler}, \&
  {Y{\i}ld{\i}z}}]{vanDishoeck11}
{van Dishoeck}, E.~F., {Kristensen}, L.~E., {Benz}, A.~O., {et~al.} 2011,
  \pasp, 123, 138

\bibitem[{{Vassilev} {et~al.}(2008){Vassilev}, {Meledin}, {Lapkin}, {Belitsky},
  {Nystr{\"o}m}, {Henke}, {Pavolotsky}, {Monje}, {Risacher}, {Olberg},
  {Strandberg}, {Sundin}, {Fredrixon}, {Ferm}, {Desmaris}, {Dochev},
  {Pantaleev}, {Bergman}, \& {Olofsson}}]{Vassilev08}
{Vassilev}, V., {Meledin}, D., {Lapkin}, I., {et~al.} 2008, \aap, 490, 1157

\end{thebibliography}
\bibliographystyle{aa}

\end{document}